\documentclass[preprint,aps,12pt,showpacs,nofootinbib,tightenlines,amsmath,amssymb]{revtex4}
\usepackage{amssymb}

\usepackage{graphicx}
\usepackage{bbm}

\def\be{\begin{equation}}
\def\ee{\end{equation}}
\def\bea{\begin{eqnarray}}
\def\eea{\end{eqnarray}}
\def\bsp{\be\begin{split}}
\def\la{\langle}
\def\ra{\rangle}
\def\dag{\dagger}
\def\bes{\be  \begin{split}}

\def\p{\partial}
\def\la{\langle}
\def\ra{\rangle}
\def\ads{AdS/CFT}

\def \vu{v_u}
\def \vs{v_s}
 \makeatletter

\newcommand{\Rmnum}[1]{\expandafter\@slowromancap\romannumeral #1@}
\makeatother
\begin{document}

\title{Predictive AdS/QCD Model \\ for Mass Spectra of Mesons with Three Flavors}
\author{Yan-Qin Sui$^{\dagger}$, Yue-Liang Wu$^{\dagger}$ and Yi-Bo Yang$^{\dagger \ddagger}$}
\affiliation{$^{\dagger}$ Kavli Institute for Theoretical Physics China (KITPC)
\\ Key Laboratory of Frontiers in Theoretical Physics
\\  Institute of Theoretical Physics\,\, Chinese Academy of Sciences
\\$^{\ddagger}$ Institute of High Energy Physics\,\, Chinese Academy of Sciences
\\  Beijing, 100190, P.R.China }

\email{sehemi@itp.ac.cn, ylwu@itp.ac.cn, yangyb@ihep.ac.cn}
 \date{\today}


\begin{abstract}
The predictive soft-wall AdS/QCD model with a modified 5D metric at the infrared region is constructed to obtain a non-trivial dilaton solution for three flavor quarks $u$, $d$ and $s$. Such a model is shown to incorporate both the chiral symmetry breaking and linear confinement. After considering some high-order terms including the $U(1)_L\times U(1)_R$ chiral symmetry breaking term, we find that the resulting predictions for the SU(3) octet and singlet resonance states of pseudoscalar, scalar, vector and axial-vector mesons agree well with the experimentally confirmed resonance states. Contributions from the instanton effects given by the determinant term are also discussed. It is observed that the chiral symmetry breaking phenomena of $SU(3)_L\times SU(3)_R$ and $U(1)_L\times U(1)_R$ can be well described in this model, while the SU(3) flavor symmetry breaking effect due to quark mass difference in the source term is not enough to explain all of the current experimental data.
\end{abstract}

\maketitle


\section{Introduction}

Strong interactions of quarks are described in the standard model
(SM) by an $SU(3)$ gauge theory known as quantum chromodynamics
(QCD) \cite{Fritzsch:1973pi}. As the gauge group is non-Abelian, the
gluons have direct self-interactions that lead to the well-known
asymptotic freedom \cite{Gross:1973id,Politzer:1973fx} due to a
negative beta function $\beta(\mu)$, which causes the coupling
constant $\alpha_{s}(\mu)$ decreasing at short distances or Ultraviolet(UV)
region, so that perturbative QCD at the UV region works well. At
low energies or Infrared(IR) region, perturbative methods are no longer
applicable as the coupling constant $\alpha_{s}(\mu)$ grows in the
IR region. We are currently unable to solve from first principle the low
energy dynamics of QCD, one can then construct effective quantum
field theories to describe the low energy features of QCD, such as
dynamically generated spontaneous symmetry breaking
\cite{Nambu:1960xd}. It has been shown in ref.~\cite{DW} that such a
dynamically generated spontaneous chiral symmetry breaking can lead
to the consistent mass spectra for both the lowest lying nonet
pseudoscalar mesons and nonet scalar mesons. Though the resulting
mass spectra for the ground states were found to agree well with the
experimental data, it is not manifest in a chiral effective field
theory how to characterize the excited meson states.

The Anti-de Sitter/Conformal Field Theory (AdS/CFT) duality
conjectured by Maldacena \cite{Maldacena:1997re} and further
developed in \cite{Gubser:1998bc,Witten:1998qj} has shed new light
on solving the problem of strongly coupled gauge theories.
Subsequently, this correspondence is aggressively expanded to
include QCD. There are two types of AdS/QCD models, one is called hard-wall AdS/QCD
\cite{Erlich:2005qh,Sakai:2004cn,Shock:2006qy,SWWX,Wu:2007tza,Damgaard:2008zs,Zhang:2010sb}, the other is
known as a soft-wall AdS/QCD \cite{Karch:2006pv,Gherghetta:2009ac,Colangelo:2008us,Sui:2009xe,Iatrakis:2010zf,Vega:2010ne,Branz:2010ub,Jin:2010xf,Iatrakis:2010jb,Vega:2010ry}.
There is also another interesting way to calculate the mass spectra for light mesons and baryons by using the approach of Light-Front holography\cite{deTeramond:2009xk,Brodsky:2009bd,Brodsky:2010kn,deTeramond:2010we,Brodsky:2010cq,Brodsky:2010px}.

The soft wall AdS/QCD model has been applied to characterize the basic trend of excited states by introducing the dilaton as a special background field. The 5D action of soft wall AdS/QCD model can be
written as follows
 \be\label{lagran1}
 S_5=\int
d^5x\sqrt{g}e^{-\Phi(z)}\,{\rm{Tr}}\left[|DX|^2-m_X^2|X|^2- \lambda
|X|^4 - \frac1{4g_5^2}\left(F_L^2+F_R^2\right)  \right]
 \ee
with $\Phi(z)\sim z^2$ playing the role of a soft cut-off. Where $g=|\det g_{MN}|$, $X(x,z)\equiv [\la X(z)\ra+S(x,z)]e^{2i\pi(x,z)}$, $D^MX=\p^MX-i A_L^MX+i X A_R^M$, and $F_{\mu\nu} = \p_\mu A_\nu -\p_\nu A_\mu - i[A_\mu, A_\nu]$.
Here $S(x,z)=S^a(x,z) t^a$,  $\pi(x,z)=\pi^a(x,z) t^a$, $V^M=V^{M~a}t^a$ and $A^M=A^{M~a}t^a$ (or $A_{L}^M= V^M - A^M$ and $A_{R}^M= V^M + A^M$, with ${\rm{Tr}}[t^at^b]=\delta^{ab}/2$) correspond to the scalar, pseudoscalar, vector and axial-vector meson fields. Here, the term $\lambda|X|^4$ was introduced to improve the masses of scalar mesons which we have
discussed in our previous paper \cite{Sui:2009xe}.  $X(z)$ is the solution of the minimal condition in 5D space, it  has the following form for the case of three flavors
    \begin{displaymath}
     X(z) =\frac{1}{2}\left( \begin{array}{ccc}
                      \vu(z) & 0&0 \\
                      0 & \vu(z)&0 \\
                      0 &0  &\vs(z)\\
                   \end{array} \right).
    \end{displaymath}

For simplicity, SU(2) symmetry remains to be considered as a good symmetry. The parameter $g_5$ is found to be $g_5^2 = 12\pi^2/N_c$
\cite{Erlich:2005qh} with $N_c$ being the color number, and $m_X^2=-3$ is fixed by \ads ~ correspondence.

In our previous paper \cite{Sui:2009xe}, we have constructed a predictive soft-wall AdS/QCD model by
simply modifying the background metric at the infrared region, which can result in a non-trivial dilaton solution. Such a model has been shown to incorporate both the chiral symmetry breaking and linear confinement, and lead to a consistent prediction for all of the resonance scalar, pseudoscalar, vector and axial vector mesons with agreement better than $10\%$ in comparison with the experiment data. Here we shall extend such a predictive AdS/CD  model to include three flavor quarks.

The problem and challenge of including three flavor quarks mainly come from the SU(3) flavor symmetry breaking due to the difference between strange quark and up/down quark masses, which leads to the different values $v_u(z)$ and $v_s(z)$ in AdS/QCD models. In general, by considering the effects of SU(3) flavor symmetry breaking, we shall be able to make a prediction for the mass spectra of all SU(3) octet and singlet resonance mesons. The paper is organized as follows: In section \ref{sec:includeS}, we shall first apply the same dilaton solution obtained in\cite{Sui:2009xe} with appropriate parameters concerning the strange quark to present an intuitive prediction for the SU(3) octet and singlet meson mass spectra. In section \ref{sec:add}, we will consider several physically meaningful higher-order interaction terms to improve the predictions for the mass spectra of mesons, which includes the $U(1)_L \times U(1)_R$ symmetry breaking term. By appropriately solving the minimal conditions including the SU(3) symmetry breaking effects, we arrive at a more reasonable prediction for the mass spectra of all SU(3) octet and singlet ground-state and resonance mesons. The possible contributions from the determinant term due to the instanton effects are discussed in section \ref{sec:determinant}. In section \ref{sec:mixing}, we discuss possible effects from the mixing term between isosinglet and singlet mesons. Our conclusions and remarks are presented in the last section.

\section{Predictive AdS/QCD Model With Three Flavor Quarks}\label{sec:includeS}

Here we take the same metric as the one introduced in \cite{Sui:2009xe}
 \be \label{metric}
ds^2=a^2(z)\left(\eta_{\mu\nu}dx^\mu dx^\nu-dz^2\right);\qquad
a^2(z)=(1+\mu_g^2\,z^2)/z^2
 \ee
where $\eta_{\mu\nu}=\rm{diag}\left(1,-1,-1,-1\right)$, and $\mu_g$
is a constant mass scale.

As we have shown in \cite{Sui:2009xe} that the background dilaton $\Phi$  has an asymptotic behavior
\be\label{dilation0}
 \Phi(z\to\infty)=\mu_d^2 \,z^2
 \ee
where the parameter $\mu_d$ sets the meson mass scale, which actually relates to the mass scale $\mu_g$ when the $X(z)$ is known.

Considering the independent variations corresponding to $\vu$ and $v_s$, we arrive at the following two minimal conditions:
\bea\label{xexpect}
 \p_z\left(a(z)^3e^{-\Phi(z)}\p_z\vu(z)
 \right)-a(z)^5e^{-\Phi(z)}m_X^2\vu(z)-\frac{\lambda}{2}a^5(z)e^{-\Phi(z)}\vu(z)^3=0.\nonumber\\
 \p_z\left(a(z)^3e^{-\Phi(z)}\p_z v_s(z)
 \right)-a(z)^5e^{-\Phi(z)}m_X^2 v_s(z)-\frac{\lambda}{2}a^5(z)e^{-\Phi(z)} v_s(z)^3=0.
 \eea

In the chiral limit or with exact SU(3) symmetry, one has $v_s(z)$=$v_u(z)$, then the above two equations are reduced to one, so that the dilaton solution obtained from $v_u(z)$ in \cite{Sui:2009xe} can directly be applied. As shown in \cite{Sui:2009xe}, the form of $\vu$ in model IIb can lead to the better mass spectra for resonance mesons in comparison with experimental data.  Thus, we shall take the similar form for $\vu(z)$ and $v_s(z)$ but with SU(3) symmetry breaking effects:
\be
\vu(z)= z(m_u\zeta+\frac{\sigma_u }{\zeta}z^2)(1+f_u z^4)^{-5/8}, \
v_s(z)= z(m_s\zeta+\frac{\sigma_s }{\zeta}z^2)(1+f_s z^4)^{-5/8}.
\ee
where $m_u$ and $\sigma_u$ are interpreted via AdS/CFT duality as the
up quark or down quark mass and quark condensate, respectively.  Similarly
$m_s$ and $\sigma_s$ correspond to the strange quark mass and its condensate. The normalization $\zeta$ is fixed by QCD with
$\zeta=\sqrt{3}/(2\pi)$ \cite{Damgaard:2008zs}. With the above given forms for $v_u(z)$ and $v_s(z)$, we have the relation for the two mass scales $\mu_d^2 = 3 \mu_g^2$.  The parameters $m_u$, $\sigma_u$, $m_s$ and $\sigma_s$ are determined by using the well measured pion and kaon meson masses $m_\pi =139.6$ MeV, $m_K=493.7$ MeV and their decay constants $f_\pi=92.4$ MeV and $f_K=113$ MeV. The parameters $f_u$ and $f_s$ are fixed from minimizing the breaking of Gell-Mann-Oakes-Renner(GMOR) relations $f_{\pi}^2m_{\pi}^2 \simeq 2m_u\sigma_u$ and $f_{K}^2m_{K}^2 \simeq 2m_s\sigma_s$, numerically it is at a few percent level. Other parameters are fitted by optimizing the mass spectrum of all mesons. The numerical values of the input parameters with three flavor quarks are given in Table I, where the parameters concerning two flavor quarks are taken to be the same as the ones given in \cite{Sui:2009xe}. All the quoted experimental data for comparison are taken from the particle data group(PDG) \cite{Amsler:2008zzb}.
\begin{table}[ht!]
\begin{center}
\begin{tabular}{|c|c|c|c|c|c|c|c|}
\hline
 $m_u$  (MeV) & $\sigma_u^{1/3}$ (MeV)  &  $m_s$  (MeV) & $\sigma_s^{1/3}$ (MeV) &$\mu_d$ (MeV)& $f_u$ & $f_s$& $\lambda$  \\
\hline \hline
        3.86        &  277.3      &  108.4  &  247.2  & 490&1.1&0.35&25 \\
\hline
\end{tabular}
\caption{The input parameters in the present model with three falvors.}\label{parameterK}
\end{center}
\end{table}

\subsection{Pseudoscalar Mesons}

Separating the quadratic term of the pseudoscalar field $\pi(x,z)=\pi^a(x,z) t^a$ from the action in Eq.\,(\ref{lagran1}), and decomposing the axial-vector field in terms of its transverse and longitudinal components ($A_{\mu}^{a}=A_{\mu \bot}^{a}+\partial_{\mu}\phi^{a} $), we can obtain the equation of motion for the SU(3) octet and singlet pseudoscalar mesons:
 \be\label{pseudoscalareom01}
\p_z[a(z)^3e^{-\Phi}(M_V^2+M_A^2)_{ab}\p_z\pi_n^b]+a(z)^3e^{-\Phi}m_{\pi_n}^{a}[(M_V^2+M_A^2)_{ab}\pi_n^b-(M_A^2)_{ab}\phi^b_n]=0
   \ee
 \be\label{pseudoscalareom02}
\p_z[a(z)e^{-\Phi}\p_z\phi^a_n]+g_5^2a(z)^3e^{-\Phi}(M_A^2)_{ab}(\pi_n^b-\phi^b_n)=0
   \ee
 with $a,\ b= 1,2,\cdots, 9$. Where $M_V^2$ and $M_A^2$ are the $9\times 9$ matrices defined as follows
 \be\label{Vmass}
 M_V^2=\left(
             \begin{array}{cccc}
               \textbf{0}_{3\times 3} & 0 & 0 & 0 \\
               0 & \frac{1}{4}(\vu(z)-v_s(z))^2\textbf{1}_{4\times 4} & 0& 0  \\
               0 & 0 & 0& 0 \\
               0 & 0 & 0& 0 \\
             \end{array}
           \right),
           \ee
  \be\label{AVmass}
   M_A^2=\left(
             \begin{array}{cccc}
               \vu(z)^2\textbf{1}_{3\times 3} & 0 & 0 & 0\\
               0 & \frac{1}{4}(\vu(z)+v_s(z))^2\textbf{1}_{4\times 4} & 0& 0 \\
               0 & 0 & \frac{1}{3}(\vu(z)^2+2v_s(z)^2)& -\frac{\sqrt{2}}{3}(v_s(z)^2-\vu(z)^2)\\
                  0 & 0 & -\frac{\sqrt{2}}{3}(v_s(z)^2-\vu(z)^2) &\frac{1}{3}(2\vu(z)^2+v_s(z)^2)\\
             \end{array}
           \right),
           \ee
The fields $\pi_n^{1,2,3}$, $\pi_n^{4,5}$, $\pi_n^{6,7}$, and $\pi_n^8$ correspond to
isovector, isodoublet, and isosinglet pseudoscalar mesons in the SU(3) octet states, and $\pi_n^9$ corresponds to the SU(3) singlet meson state. For simplicity, we may first ignore the mixing term between the isosinglet and singlet meson states, which is corresponding to
$g^{\mu\nu}(\p_\mu\pi^8-A_\mu^8)(\p_\mu\pi^9-A_\nu^9)$, so that $m_{\pi_n}^{ab}$ can be replaced, in a good approximation, by the diagonal mass matrix $m_{\pi_n^a}^2$.

The above equation Eqs.\,(\ref{pseudoscalareom01}) and Eq.\,(\ref{pseudoscalareom02}) can be solved
by the shooting method with the following boundary conditions:

\be
\pi(z\to 0) =0, \quad \partial_z\pi(z\to \infty) =0; \quad \phi(z\to 0) =0,\quad \partial_z\phi(z\to \infty) =0.
\ee

With a positive dilaton solution and the forms of $v_u(z), v_s(z), a(z)$ given above, the result is
not sensitive to the details on the asymptotic behavior of $\pi(z)$ or $\phi(z)$ around origin. While the relation between $\partial_z\pi$ and $\partial_z\phi$ around origin becomes important, which can be obtained from Eq.\,(\ref{pseudoscalareom01}) and Eq.\,(\ref{pseudoscalareom02}) to be
\bea
& & a(z)\big(g_5^2a(z)^2(M_V^2+M_A^2)_{aa}\p_z\pi_n^a-m_{\pi_n^a}^2\p_z\phi^a_n\big)|_{z\rightarrow0} = \\
& & \int^{\infty}_{0}\text{d}z \big(e^{-\Phi}m_{\pi_n^a}^2g_5^2 a(z)^3(M_V^2)_{aa}\pi_n^a) +e^{-\Phi}a(z)\big(g_5^2a(z)^2(M_V^2+M_A^2)_{aa}\p_z\pi_n^a-m_{\pi_n^a}^2\p_z\phi^a_n\big)|_{z\rightarrow\infty}. \nonumber
\eea

For a positive dilaton solution, the integration on the right-hand side of the above equation is finite, which gives a definitive relation for $\partial_z\pi$ and $\partial_z\phi$ around origin. The numerical results are given in Table\,\ref{pseudoscalarmassesk}

 \begin{table}[ht!]
\begin{center}
\begin{tabular}{|c|c|c|c|c|c|c|c|c|}
\hline
        n &$\pi$ exp.(MeV)  &Theory& $K$ exp.(MeV)   &Theory& $\eta$ exp.(MeV)&Theory& $\eta'$ exp.(MeV)&Theory   \\
\hline\hline
       0  &   139.6           & 139.6                          & 493.7            &   493.7         &  $547.853\pm0.024$     & 528.2 & $957.78 \pm 0.06$  &460 \\
\hline
         1 &$1350 \pm 100$   & 1490                      &1460              &1530            &$1476\pm4$                   &1546            &---            &1523 \\
\hline
         2 &$1816 \pm 14$     &1733                       & 1830               &  1769   & $1756\pm9$                       & 1783              & ---        &1763\\
\hline
        3  &---&1933        &---&1966             &---&1979               &---&1961           \\
\hline
        4&---&2103           &---& 2134            &---&2147               &---&2129           \\
\hline
        5&---&2251             &---& 2281            &---&2293               &---&2276           \\
\hline
\end{tabular}
\caption{The predicted mass spectra for
pseudoscalar mesons in comparison with experiment.} \label{pseudoscalarmassesk}
\end{center}
\end{table}

From Table\,(\ref{pseudoscalarmassesk}), it is seen that the resulting meson mass spectra agree well with the data except for the singlet pseudoscalar $\eta'$ which will be discussed in detail below.

Before proceeding, we would like to address that in our previous paper\cite{Sui:2009xe} we have adopted two different ways to carry out the calculations for the mass spectra of pseudoscalar mesons. One is the way used above with solving two coupled equations by the shooting method. Another way is to first eliminate the longitudinal component field $\phi$ from the two coupled equations and obtain a single equation for the $\pi$ field. In the case with two flavor and SU(2) symmetry, the single equation can be written as follows
 \be \label{pit}
 -\partial_z^2 \tilde{\pi}(q,z)+ V_{\pi}(z)~\tilde{\pi}(q,z)= m_{\pi_n}^2\tilde{\pi}(q,z),
 \ee
 with the definitions $\tilde{\pi}(q,z)\equiv \partial_z \pi(q,z)$ and
 \be  \label{piv}
 \begin{split}
 V_{\pi}(z)&=g_5^2
 a(z)^2v(z)^2+\frac{\Phi'^2+2\Phi''}{4}+\frac{15a'(z)^2}{4a(z)^2}-\frac{3a'(z)(v(z)\Phi'-2v'(z))}{2a(z)v(z)}\\
 & -\frac{3a''(z)}{2a(z)} +\frac{2v'(z)^2}{v(z)^2}-\frac{\Phi'v'(z)+v''(z)}{v(z)}.
 \end{split}
 \ee
which can be solved by the shooting method with the boundary conditions: $\tilde{\pi}(z\to 0) =0$ and $\partial_z \tilde{\pi}(z\to \infty) =0$. To be more explicit, we present the numerical results obtained by two ways in the Table \ref{check}.  As a consequence, we arrive at the same results by two different ways.
\begin{table}[ht!]
\begin{center}
\begin{tabular}{|c||c|c|c|c|c|c|c|}
\hline
       $m_\pi$(MeV)&n=0&n=1&n=2&n=3&n=4&n=5& n=6\\
\hline
experiment   & 139.6&$1350 \pm 100$&$1816 \pm 14$&-- &--&--&-- \\
\hline
$\pi(q,z)$      & 139.6&1474&1733 &1956 &2155&2336 &2503\\
\hline
$\tilde{\pi}(q,z)\equiv \partial_z\pi(q,z)$  & 139.6&1474&1733 &1956 &2155&2336 &2503\\
\hline
\end{tabular}
\caption{The mass spectra of pseudo-scalar mesons with two different ways.} \label{check}
\end{center}
\end{table}

\subsection{Scalar Mesons}

The equation of motion for the scalars is the same as the one discussed in \cite{Sui:2009xe}.
Separating the quadratic term of the scalar field $S(x,z)=S^a(x,z) t^a$ from the action in Eq.\,(\ref{lagran1}),
and assuming the decomposition $S^a(x,z)=\sum_{n}\mathcal S^a_n(x)S^a_n(z)$ with defining $ S_n(z)\equiv
e^{\omega_s/2}s_n(z)=e^{(\Phi-3\log a(z) )/2}s_n(z)$, we arrive at the following equation of motion(EOM):
    \be\label{scalareom1}
  -\p_z^2s_n^a(z)+\left(\frac1{4}\omega'^2-\frac1{2}\omega''+a(z)^2m_X^2\right) s_n^a +\lambda a(z)^2(M_S^2)_{ab} s_n^b(z)=m_{S_n^a}^2s_n^a(z)
   \ee
   with %
  \be
  M_S^2=\left(
             \begin{array}{cccc}
              \frac{3\vu(z)^2}{2} \textbf{1}_{3\times 3} & 0 & 0& 0  \\
               0 & \frac{\vu(z)^2+\vu(z)v_s(z)+v_s(z)^2}{2}\textbf{1}_{4\times 4} & 0& 0  \\
               0 & 0 & \frac{\vu(z)^2+2v_s(z)^2}{2}& -\frac{v_s(z)^2-\vu(z)^2}{2\sqrt{2}} \\
                0 & 0 & -\frac{v_s(z)^2-\vu(z)^2}{2\sqrt{2}}& \frac{2\vu(z)^2+v_s(z)^2}{2} \\
             \end{array}
           \right).
  \ee
where the fields $S_n^{1,2,3}$, $S_n^{4,5}$, $S_n^{6,7}$ and $S_n^8$ are the isovector, isodoublet, isosinglet scalar mesons in
the SU(3) octet states, and $S_n^9$ is the SU(3) singlet meson state. Again for simplicity, we shall first ignore the mixing effect between isosinglet and singlet mesons, which will be discussed in section \ref{sec:mixing}.

Adopting the shooting method to solve Eq.\,(\ref{scalareom1}) with the
boundary conditions $s_n(z\to 0) =0$ and $\partial_z s_n(z\to \infty)
=0$, we obtain the mass spectra for the scalar mesons given in Table\, \ref{scalarmassesk}.
 \begin{table}[ht!]
\begin{center}
\begin{tabular}{|c|c|c|c|c|c|c|c|c|}
\hline
        n &$a_0$ exp.(MeV)  &Theory& $K_0^*$ exp.(MeV)   &Theory & $f_0$ exp.(MeV)& Theory& $f_0$ exp.(MeV)& Theory   \\
\hline\hline
       0  &  $980 \pm 20$    &304                         & $672\pm40$                &   398    &  $980 \pm 10$     & 433 & $550^{+250}_{-150}$  &374 \\
\hline
         1 & $1474 \pm 19$   &1475                     & $1425 \pm 50$               &1497      &   $1505 \pm 6$    &1506 & $1350 \pm 150$    &1491  \\
\hline
         2 &  ---                     &1719                          & --- &  1739       &  ---                    & 1748&  $1793 \pm 7$      & 1733 \\
\hline
         3 &  ---                      &1919                           & $1945\pm 10\pm20$                              &  1939    &     $2103 \pm 8$  &1947&  $1992 \pm 16$    & 1933\\
\hline
         4 &---                        &2090                           & ---                                &   2109   &  $2337 \pm 14$   &2117& $2189 \pm 13$       & 2103\\
\hline
        5&---&  2237      &---& 2256            &---&2264               &---&  2251            \\
\hline
\end{tabular}
\caption{The predicted mass spectra for scalar mesons in comparison with experiment.} \label{scalarmassesk}
\end{center}
\end{table}

From the Table\,\ref{scalarmassesk}, it is easily seen that the resulting excited resonance
meson states agree well with the data, while the masses for the SU(3) octet ground states are much smaller than the experimental data, especially, the SU(3) flavor symmetry breaking effects are not big enough to explain all of the data.

\subsection{Vector Mesons}

The equation of motion for the vector meson field is:
  \be\label{vectorom1}
  -\p_z^2V_n^a+\omega'\p_zV_n^a+g_5^2a(z)^2(M_V^2)_{aa}V_n^a=m^2_{V_n^a}V_n^a,
  \ee
where the matrix $M_V^2$ is diagonal and given in eq.(\ref{Vmass}).  The fields $V_n^{1,2,3}$, $V_n^{4,5}$, $V_n^{6,7}$ and $V_n^8$ correspond to the isovector, isodoublet and isosinglet vector mesons in the SU(3) octet states, and $V_n^9$ corresponds to the SU(3) singlet vector meson state.

With the definition $V_n\equiv e^{\omega/2} v_n=e^{\left(\Phi(z)-\log
a(z)\right)/2}v_n$, the above equation of motion can be rewritten as
 \be
 -\p_z^2v_n^a+\left(\frac1{4}\omega'^2-\frac1{2}\omega''+g_5^2a(z)^2(M_V^2)_{aa}\right)v_n^a=m^2_{V_n^a}v_n^a.
 \ee
which can be solved by the shooting method. Using the boundary conditions $v_n(z\to 0) =0$ and $\partial_z
v_n(z\to \infty) =0$, we obtain the vector meson mass spectra which are presented in
Table\,\ref{vectormassesk}.
\begin{table}[ht!]
\begin{center}
\begin{tabular}{|c|c|c|c|c|c|c|c|c|}
\hline
        n &$\rho$ exp.(MeV)  &Theory & $K^*$ exp.(MeV)   &Theory & $\phi$ exp.(MeV)&Theory & $\omega$ exp.(MeV)& Theory   \\
\hline\hline
         0 &  $775.5 \pm 1$   &750  &  $891.66 \pm 0.26$     &755 &$1019.455 \pm 0.020 $   &750 & $782.65 \pm 0.12$   &750      \\
\hline
         1 &  $1465 \pm 25$  &1491   &  $1414 \pm 15$  & 1493 & $1680 \pm 20$                 &1491  &$1400-1450$              &1491      \\
\hline
         2 & $1720 \pm 20$  & 1745   & $1717 \pm 27$    & 1747  & $2175 \pm 15$               & 1745 &$1670 \pm 30$         &1745      \\
\hline
         3 & $1909 \pm 30$     &1945  & ---                      &  1947  & ---                                  &  1945& ---                             & 1945  \\
\hline
         4 & $2149 \pm 17$   & 2114   & ---                        & 2116   & ---                                  &2114& ---                           & 2114   \\
\hline
         5 & $2265 \pm 40$   &  2259   & ---                      &   2261   & ---                                    &  2259 & ---                         &  2259    \\
\hline
\end{tabular}
\caption{The predicted mass spectra for vector mesons in comparison with experiment.} \label{vectormassesk}
\end{center}
\end{table}

It is seen from Table\, \ref{vectormassesk} that the resulting predictions for the $\rho $, $K^* $, $\omega$ vector mesons and their excited states agree well with the data. While the prediction for the isosinglet vector meson mass remains smaller than the data. In fact, the isovector, isosinglet and singlet vector mesons have the same EOM as given in Eq.\,(\ref{vectorom1}).

\subsection{Axial Vector Mesons}

From the action Eq.\,(\ref{lagran1}) with the gauge condition $A_5=0$, one can
derive the equation of motion for the perpendicular component of axial-vector
field as follows
\be\label{Atransverse}
e^{\Phi}\partial_z(a(z)e^{-\Phi}\partial_zA_{n}^a)+a(z)q^2A_{n}^a-a(z)^3g_5^2(M_A^2)_{ab}A_{n}^b=0
\ee
With the redefinition $A_n^a\equiv e^{\omega/2} a_n^a=e^{\left(\Phi(z)-\log
a(z)\right)/2}a_n^a$, the above equation of motion can be
reexpressed as
  \be\label{axialeom}
 -\p_z^2a_n^a+\left(\frac1{4}\omega'^2-\frac1{2}\omega''\right) a_n^a +g_5^2(z)a^2(M_A^2)_{ab}a_n^b=m^2_{A_n^a}a_n^a.
  \ee
where the matrix $M_A^2$ is given in eq.(\ref{AVmass}). The fields $A_n^{1,2,3}$, $A_n^{4,5}$, $A_n^{6,7}$ and $A_n^8$ are isovector, isodoublet and isosinglet axial-vector mesons in the SU(3) octet states, and $A_n^9$ is the SU(3) singlet axial-vector meson state. As a good approximation, we first ignore the mixing term between the isosinglet and singlet states, which will be discussed in section IV.

With the boundary conditions $a_n(z\to 0) =0$ and $\partial_z a_n(z\to
\infty) =0$, the resulting mass spectra by using the shooting method
is given in Table\,\ref{axial-vectormassesk}.
 \begin{table}[ht!]
\begin{center}
\begin{tabular}{|c|c|c|c|c|c|c|c|c|}
\hline
        n &$a_1$ exp.(MeV)  &Theory & $K_1$ exp.(MeV)   &Theory & $f_1$ exp.(MeV)&Theory& $f_1$ exp.(MeV)&Theory   \\
\hline\hline
       0  &   $1230 \pm 40$    &829       &  $1272 \pm 7$        & 870 &$1426.4 \pm 0.9$    & 890  &$1281.8 \pm 0.6$ &860\\
\hline
         1 &  $1647 \pm 22$     &1531     &  $1403 \pm 7$       &1553     & ---                          &1563  & $1518 \pm 5$   &1547 \\
\hline
         2 & $1930^{+30}_{-70}$  &1783   & $1650 \pm 50$    & 1803   &---                           & 1812&  ---                 & 1797 \\
\hline
         3 & $2096 \pm 122$      & 1982    &---                         &  2001    &    ---                         &2010& ---                  & 1996\\
\hline
         4 & $2270^{+55}_{-40}$    &2150  & ---                       &   2169   & ---                           &2178&---                        & 2164\\
\hline
         5&---&  2296             &---&2315             &---& 2323          &---&  2310          \\
\hline
\end{tabular}
\caption{The predicted mass spectra for axial-vector mesons in comparison with experiment.} \label{axial-vectormassesk}
\end{center}
\end{table}

It is seen from Table\,\ref{axial-vectormassesk} that the resulting excited resonance states for the $a_1$ mesons agree well with the data, while the ground state mass remains much below to the data. The predicted mass spectra for $K_1$ and $f_1$ mesons are very similar to the ones for $a_1$ mesons, their departure to the experimental data is more than $10\%$. We shall improve the above prediction in next section.

\section{High Order Term Corrections to Mass Spectra}\label{sec:add}

It has been shown in Table\,\ref{pseudoscalarmassesk} that the mass of singlet pseudoscalar meson $\eta'$ is much below to the experimental data. To understand such a big discrepancy, we shall consider an additional $U(1)_L \times U(1)_R$ symmetry breaking term at high order with the explicit form $(Tr\ln X-Tr\ln X^\dag)^2 Tr(XX^\dag)$ which is motivated from the large N chiral dynamics \cite{Witten}.

Also the masses for the ground-state axial-vector mesons given in Table\, \ref{axial-vectormassesk} are much smaller than the data. To improve the prediction, we shall add a high order term  $ i (D_M X^\dag D_N X-D_N X^\dag D_M X)A^{NM}$($N, M=1,5$) into the action. Note that similar terms $iA_{\mu\nu}(D^{\mu}U^{+}D^{\nu}U-D^{\nu}U^{+}D^{\mu}U)$ and $iV_{\mu\nu}(D^{\mu}U^{+}D^{\nu}U+D^{\nu}U^{+}D^{\mu}U)$($\mu,\nu=1,4$) with $V_{\mu\nu}=D_{\mu}v_{\nu}-D_{\nu}v_{\mu},\ A_{\mu\nu}=D_{\mu}a_{\nu}-D_{\nu}a_{\mu}$ actually appear as the $p^4$ order terms via the momentum expansion in the chiral effective field theory.

With the above considerations, the modified effective action with relevant high order terms is found to be:
 \bea\label{lagran2}
S_5&=&\int d^5x\sqrt{g}e^{-\Phi(z)}\,{\rm{Tr}}\{|DX|^2-m_X^2|X|^2-
\lambda |X|^4-\frac1{4g_5^2}(F_L^2+F_R^2) \}\nonumber\\
&+&  c_a\ i(D_M X^\dag D_N X-D_N X^\dag D_M X)A^{NM}+ c_{1}({\rm{Tr}}\ln X-{\rm{Tr}}\ln X^\dag)^2{\rm{Tr}}XX^\dag
 \eea
It is noticed that the additional two terms don't change the minimal conditions, so that we can take the same dilaton solution as the one obtained in previous section.

The interaction term $i(D_M X^\dag D_N X-D_N X^\dag D_M X)A^{NM}$ only influences the equation of motion for axial-vector mesons, its coupling coefficient can be determined by the ground state mass of axial-vector meson. The numerical value is found to be $c_a=12$. The chiral U(1) symmetry breaking term $(Tr\ln X-Tr\ln X^\dag)^2 Tr(XX^\dag)$ only changes the equation of motion for the singlet pseudoscalar meson $\eta'$, its coupling coefficient is determined to be $c_{1}=0.38$ from the mass of $\eta'$.
Other parameters are taken to be the same values as the ones given in Table\, \ref{parameterK}. Note that these two terms will not change the mass spectra for scalar and vector mesons. Our improved predictions for the pseudoscalar and axial-vector mesons are discussed in detail below.

\subsection{Pseudoscalar Mesons}

The equation of motion for the SU(3) octet pseudoscalar mesons is the same as the one given in previous section. For the singlet of pseudoscalar meson $\eta'$, the equation of motion is modified to be
 \bea\label{pseudoscalareom09}
&&\p_z[a(z)^3e^{-\Phi}(M_V^2+M_A^2)_{99}\p_z\pi_n^9]
+a(z)^3e^{-\Phi}m_{\eta'_n}^{2}[(M_V^2+M_A^2)_{99}\pi_n^9-(M_A^2)_{99}\phi_n^9]\nonumber\\
&&- 8c_{1}a(z)^5e^{-\Phi}(2\vu(z)^2+v_s(z)^2)\pi_n^9=0
   \eea
where the mixing term is ignored in a good approximation, which will be discussed in next section.

Taking the same boundary conditions given in the previous section, we obtain the improved prediction for the $\eta'$ meson mass. For completeness, we present the results in Table\, \ref{pseudoscalarmassesk2}:

 \begin{table}[ht!]
\begin{center}
\begin{tabular}{|c|c|c|c|c|c|c|c|c|}
\hline
        n &$\pi$ exp.(MeV)  &Theory& $K$ exp.(MeV)   &Theory& $\eta$ exp.(MeV)&Theory& $\eta'$ exp.(MeV)&Theory   \\
\hline\hline
       0  &   139.6           & 139.6                          & 493.7            &   493.7         &  $547.853\pm0.024$     & 528.2 & $957.78 \pm 0.06$  &957.9 \\
\hline
         1 &$1350 \pm 100$   & 1490                      &1460              &1530            &$1476\pm4$                   &1546            &---            &1584\\
\hline
         2 &$1816 \pm 14$     &1733                       & 1830               &  1769   & $1756\pm9$                       & 1783              & ---        &1814\\
\hline
        3  &---&1933        &---&1966             &---& 1979              &---& 2005         \\
\hline
        4&---&2103           &---& 2134            &---& 2147              &---&  2169        \\
\hline
        5&---&2251             &---& 2281            &---&2293               &---& 2313         \\
\hline
\end{tabular}
\caption{The predicted mass spectra for pseudoscalar mesons with including chiral U(1) symmetry breaking term.} \label{pseudoscalarmassesk2}
\end{center}
\end{table}

\subsection{Axial Vector Mesons}

As the term $i (D_M X^\dag D_N X-D_N X^\dag D_M X)A^{NM}$ influences the equation of motion for the axial-vector mesons by modifying the matrix $M_A^2$, which has the following form
 \be
 \hat{M}_A^2=\left(
             \begin{array}{cccc}
             F_1(z)\textbf{1}_{3\times 3} & 0 & 0 & 0\\
               0 &
               F_2(z)\textbf{1}_{4\times 4}
               & 0& 0 \\
               0 & 0 &F_3(z)& F_{89}(z)\\
                  0 & 0 &F_{89}(z)&F_4(z)\\
             \end{array}
           \right),
           \ee

with
 \be
 F_1(z)=\vu(z)^2-c_1\frac{a(z)'}{a(z)^3}\vu(z)\vu(z)'
 \ee
 \be
 F_2(z)=\frac{1}{4}(\vu(z)+v_s(z))^2-c_1\frac{a(z)'}{a(z)^3}\frac{(\vu(z)+v_s(z))(\vu(z)'+v_s(z)')}{4}
 \ee
 \be
F_3(z)=\frac{1}{3}(\vu(z)^2+2v_s(z)^2)-c_1\frac{a(z)'}{a(z)^3}\frac{\vu(z)\vu(z)'+2v_s(z)v_s(z)'}{3}
 \ee
  \be
F_4(z)=\frac{1}{3}(2\vu(z)^2+v_s(z)^2)-c_1\frac{a(z)'}{a(z)^3}\frac{2\vu(z)\vu(z)'+v_s(z)v_s(z)'}{3}
 \ee
\be
 F_{89}(z)=-\frac{\sqrt{2}}{3}(v_s(z)^2-\vu(z)^2)+c_1\frac{a(z)'}{a(z)^3}\frac{\sqrt{2}}{3}(v_s(z)v_s(z)'-\vu(z)\vu(z)')
\ee
which only enters into the equation of motion for the axial-vector mesons, the coefficient $c_a$ is determined to be $c_a=10$ from the ground state mass of the axial-vector mesons. Taking other parameters given in Table\, \ref{parameterK} and ignoring the mixing term $F_{89}(z)$, the resulting mass spectra by using the shooting method is given in Table\, \ref{axial-vectormassesk31}.
\begin{table}[ht!]
\begin{center}
\begin{tabular}{|c|c|c|c|c|c|c|c|c|}
\hline
        n &$a_1$ exp.(MeV)  &Theory & $K_1$ exp.(MeV)   &Theory & $f_1$ exp.(MeV)& Theory & $f_1$ exp.(MeV)& Theory   \\
\hline\hline
       0  &   $1230 \pm 40$           &1204    &  $1272 \pm 7$        & 1377 &$1426.4 \pm 0.9$    &1444 &$1281.8 \pm 0.6$ &1338\\
\hline
         1 &  $1647 \pm 22$           &1608   &  $1403 \pm 7$         &1669   & ---                        &1700   & $1518 \pm 5$   &1653\\
\hline
         2 & $1930^{+30}_{-70}$    & 1838  & $1650 \pm 50$       &   1886  &---                         &  1910  &  ---                   & 1874\\
\hline
         3 & $2096 \pm 122$         & 2026  &---                              &  2068 &    ---                    &2088    & ---                      & 2057\\
\hline
         4 & $2270^{+55}_{-40}$    & 2186 & ---                           &   2224   & ---                      &2242     &---                      & 2214\\
\hline
        5&---& 2326             &---&2361             &---& 2377              &---& 2352        \\
\hline
\end{tabular}
\caption{The predicted mass spectra for
axial-vector mesons with including high order term.} \label{axial-vectormassesk31}
\end{center}
\end{table}

It is seen from the Table \ref{axial-vectormassesk31} that the improvement to the prediction for the ground states is manifest.

\section{Instanton Effect with Determinant Term}\label{sec:determinant}

It can be seen from Table\,\ref{scalarmassesk} that the prediction for the ground state masses of scalar mesons is not satisfactory, especially the mass difference between $\sigma$ and $a_0$ is opposite to the data. To improve such a situation, we may discuss the possible instanton effects by adding the determinant term of $X$ to the action:
 \bea\label{lagran3}
S_5&=&\int d^5x\sqrt{g}e^{-\Phi(z)}\,{\rm{Tr}}\{|DX|^2-m_X^2|X|^2-
\lambda |X|^4-\frac1{4g_5^2}(F_L^2+F_R^2) \}\nonumber\\
&+&c_a\ i (D_M X^\dag D_N X-D_N X^\dag D_M X)A^{NM}+ c_{1}({\rm{Tr}}\ln X-{\rm{Tr}}\ln X^\dag)^2{\rm{Tr}}XX^\dag\nonumber\\
 &-&c_0 \text{Re}(Det[X])
 \eea
we will show that for a positive $c_0$, it can really split the mass differences among $\sigma$, $a_0$ and f$_0$(980) toward the right direction.

The minimal conditions for the field X are modified to be
\bea\label{xexpect4}
V_u & \equiv& \p_z[a(z)^3e^{-\Phi(z)}\p_z\vu(z)]- e^{-\Phi(z)} a(z)^5[m_X^2\vu(z)-\frac{\lambda}{2}\vu(z)^3 -\frac{c_0}{4}\vu(z)v_s(z)] = 0 \\
V_s & \equiv & \p_z[a(z)^3e^{-\Phi(z)}\p_z v_s(z)]- e^{-\Phi(z)} a(z)^5[ m_X^2 v_s(z)-\frac{\lambda}{2} v_s(z)^3-\frac{c_0}{4}\vu(z)^2] =0.
 \eea
where $v_u$ and $v_s$ enter into both equations. To effectively find out a solution for the dilaton $\Phi(z)$, considering the combination $V_u + 2\alpha V_s$ with $\alpha$ an arbitrary parameter, then the solution is figured out by requiring the equation $V_u + 2\alpha V_s = 0$ to be insensitive to the choices for the values of the parameter $\alpha$. The dilaton solution is given by
 \bea\label{dilaton00}
  \p_z\Phi(z)&=&\Big{\{}2\{\p_z[a(z)^3\p_z\vu(z)]-a(z)^5m_X^2\vu(z)-\frac{\lambda}{2}a(z)^5\vu(z)^3 -\frac{c_0}{4}a^5(z)\vu(z)v_s(z)\} \alpha \nonumber\\
  &&+\{\p_z[a(z)^3\p_zv_s(z)]-a(z)^5m_X^2v_s(z)-\frac{\lambda}{2}a(z)^5v_s(z)^3-\frac{c_0}{4}a^5(z)\vu(z)^2\}\Big{\}}\nonumber\\
  &&/\Big{\{}a(z)^3[2\p_z\vu(z)\alpha+\p_zv_s(z)]\Big{\}}
 \eea
It can be shown that for a small $z\ll 1$, one has $\p_z\Phi(z)\propto c_0 m_u \alpha+6 \mu_g^2 z$. As the light quark mass $m_u$ is very small and $c_0$ is of order $1$, thus for a large range of $\alpha$, the $\alpha$ dependence of the dilaton solution is greatly suppressed by the mass factor $m_u$. For simplicity, choosing $\alpha=1$, the dilaton is determined when all relevant parameters are fixed.

The parameters  $m_u$, $m_s$, $\sigma_u$ and $\sigma_s$ involved in $v_u(z)$ and $v_s(z)$ are determined with the input experimental data $m_\pi =139.6$ MeV ,$m_K=493.7$ MeV, $f_\pi=92.4$ MeV, $f_K=113$ MeV.  The parameters $f_u$ and $f_s$ are fixed by minimizing the breaking of GOMR relations $f_\pi^2 m_\pi^2 \simeq m_u \sigma_u$ and $f_K^2 m_K^2 \simeq m_s \sigma_s$. The mass scale $\mu_d$ is fitted by optimizing the global behavior of mass spectra. Note that in order to ensure a large enough region for the $z^2$ power-counting of $\Phi(z)$ required for obtaining correct resonance meson states, $\sigma_s$ has to be very close to $\sigma_u$ in the present case, which results in a sizable breaking of GOMR relation for the kaon meson (up to about $30\%$). The set of parameters used in the present case is given in Table\, \ref{parameterK3}. The other parameters are taken to be the same as the ones given in previous section.
\begin{table}[ht!]
\begin{center}
\begin{tabular}{|c|c|c|c|c|c|c|c|c|c|}
\hline
 $m_u$  (MeV) & $\sigma_u^{1/3}$ (MeV)  &  $m_s$  (MeV) & $\sigma_s^{1/3}$ (MeV) &$\mu_d$ (MeV)& $f_u$& $f_s$ & $\lambda$ &$c_{1}$&$c_a$\\
\hline \hline
         3.59              &  285      & 92.8   &285&547.7& 1.1& 0.63&25&0.029&10\\
\hline
\end{tabular}
\caption{The input parameters with including the instanton effects of determinant.} \label{parameterK3}
\end{center}
\end{table}

The dilaton solution of $\Phi(z)$ with different values of $\alpha$ is plotted in Fig.1, which shows that in the physically meaningful region of $z$, $\Phi(z)$ is not sensitive to the parameter $\alpha$ in a large range of $\alpha$.

\begin{figure}[ht]
\begin{center}
\includegraphics[width=10cm,clip=true,keepaspectratio=true]{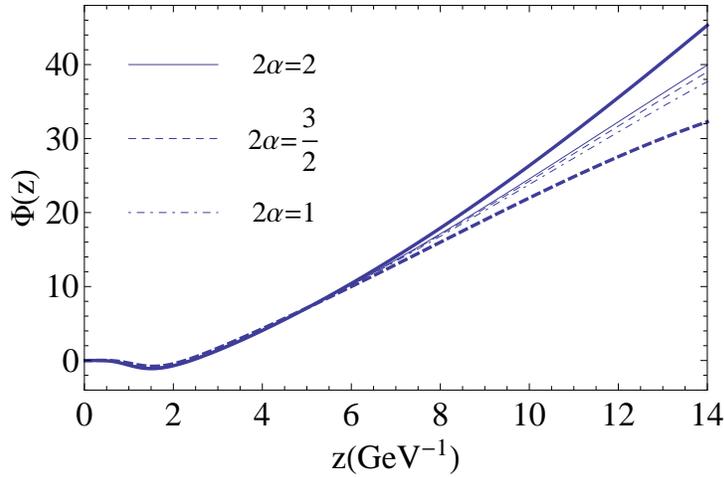}
\caption{The plot of dilaton $\Phi(z)$ for different values $\alpha$.  The thick real line is for $\alpha=\infty$ and the thick dashed line for $\alpha=0$}\label{fig:dilaton2}
\end{center}
\end{figure}

 \subsection{Pseudoscalar and Scalar Mesons}

From the above action Eq.\,(\ref{lagran3}), it is easy to see that the equation of motion for the
SU(3) octet pseudoscalar mesons is just the same as the previous ones.  While for the singlet state, its equation of motion is modified to be
 \bea\label{pseudoscalareom19}
&&\p_z\left(a(z)^3e^{-\Phi}(M_V^2+M_A^2)_{99}\p_z\pi_n^9\right)
+a(z)^3e^{-\Phi}m_{\eta'_n}^2\left((M_V^2+M_A^2)_{99}\pi_n^9-(M_A^2)_{99}\phi_n^9\right)\nonumber \\
&&- 8c_{1}a(z)^5e^{-\Phi}(2\vu(z)^2+v_s(z)^2)\pi_n^9-a(z)^5e^{-\Phi} \frac{3}{4}c_0\vu(z)^2v_s(z)\pi_n^9=0 \\
&&\p_z[a(z)e^{-\Phi}\p_z\phi_n^9]+g_5^2a(z)^3e^{-\Phi}(M_A^2)_{99}(\pi_n^9-\phi_n^9)=0
\eea
Using the same boundary conditions as the ones given in previous section, but with different input parameters given in Table. \ref{parameterK3}, we obtain the interesting results given in Table\, \ref{pseudoscalarmassesk3}.

 \begin{table}[ht!]
\begin{center}
\begin{tabular}{|c|c|c|c|c|c|c|c|c|}
\hline
        n &$\pi$ exp.(MeV)  &Theory & $K$ exp.(MeV)   &Theory & $\eta$ exp.(MeV)& Theory & $\eta'$ exp.(MeV)&Theory   \\
\hline\hline
       0  &   139.6           & 139.6                          & 493.7            &   493.7       &  $547.853\pm0.024$     & 531 & $957.78 \pm 0.06$  &957.3\\
\hline
         1 &$1350 \pm 100$   & 1566                     &1460               &1604          &$1476\pm4$               &1619        &---                      &1663 \\
\hline
         2 &$1816 \pm 14$     &1778                      & 1830               & 1813  & $1756\pm9$                       &1828            & ---               &1862\\
\hline
        3  &---&   1931             &---& 1964            &---&  1979             &---& 2017               \\
\hline
        4&---&   2072              &---& 2137           &---&    2153           &---&  2173                \\
\hline
        5&---&     2173             &---&  2250           &---&    2265           &---&  2356              \\
\hline
\end{tabular}
\caption{The predicted mass spectra for pseudoscalar mesons with including instanton effects of determinant term.} \label{pseudoscalarmassesk3}
\end{center}
\end{table}

The equation of motion for the scalar mesons are changed to be :
    \bea\label{scalareom3}
 &-&\p_z^2s_n^a(z)+[\frac1{4}\omega'^2-\frac1{2}\omega''+a(z)^2m_X^2+\lambda a(z)^2(M_S^2)_{ab}-c_0 a(z)^2(\tilde{M}_S^2)_{ab}
  ]s_n^b(z)\nonumber\\
  &=&m_{S_n^a}^2s_n^a(z)
   \eea

with the additional matrix given by
\be
\tilde{M}_S^2=\left(
       \begin{array}{cccc}
              -\frac{v_s(z)}{4} \textbf{1}_{3\times 3} & 0 & 0& 0  \\
               0 & -\frac{\vu(z)}{4}\textbf{1}_{4\times 4} & 0& 0  \\
               0 & 0 & \frac{-4\vu(z)+v_s(z)}{12}& -\frac{\vu(z)-v_s(z)}{6\sqrt{2}} \\
                0 & 0 & -\frac{\vu(z)-v_s(z)}{6\sqrt{2}}& \frac{2\vu(z)+v_s(z)}{6} \\
             \end{array}
           \right)
\ee

When ignore the mixing term between isosinglet and singlet states, which will be discussed in next section, we obtain the improved mass spectra given in Table \ref{scalarmassesk3}. Though the predicted masses for the ground states are improved, while it remains unsatisfactory to explain all of the experimental data.

 \begin{table}[ht!]
\begin{center}
\begin{tabular}{|c|c|c|c|c|c|c|c|c|}
\hline
        n &$a_0$ exp.(MeV)  &Theory & $K_0^*$ exp.(MeV)   &Theory & $f_0$ exp.(MeV)& Theory & $f_0$ exp.(MeV)&Theory   \\
\hline\hline
       0  &  $980 \pm 20$    &437      & $672\pm40$              &  510&  $980 \pm 10$     &539& $550^{+250}_{-150}$  &400\\
\hline
         1 & $1474 \pm 19$   &1551   & $1425 \pm 50$          &1574&   $1505 \pm 6$    &1584 & $1350 \pm 150$    &1552\\
\hline
         2 &  ---                    &1762    & $1945\pm 10\pm20$ & 1785&  ---                    & 1794&  $1724 \pm 7$        & 1765 \\
\hline
         3 &  ---                    &1959  &---                                & 1995&     $2103 \pm 8$  &2010&  $1992 \pm 16$      &1972\\
\hline
         4 &---                     &2129   & ---                                &  2157&  $2337 \pm 14$   &2168& $2189 \pm 13$       & 2137\\
\hline
         5&---&  2363             &---&2387             &---&  2398             &---&  2370                       \\
\hline
\end{tabular}
\caption{The predicted mass spectra for scalar mesons with including instanton effects of determinant term.} \label{scalarmassesk3}
\end{center}
\end{table}

\subsection{Vector and Axial Vector Mesons}

The equation of motion for the vector mesons and axial-vector mesons are not changed. While the input parameters are modified with the inclusion of determinant term due to the instanton effects, we then recalculate the mass spectra which are given in Table\, \ref{vectormassesk3} for vector mesons and Table\, \ref{axial-vectormassesk3} for axial-vector mesons.

\begin{table}[ht!]
\begin{center}
\begin{tabular}{|c|c|c|c|c|c|c|c|c|}
\hline
        n &$\rho$ exp.(MeV)  &Theory & $K^*$ exp.(MeV)   &Theory & $\phi$ exp.(MeV)&Theory& $\omega$ exp.(MeV)& Theory   \\
\hline\hline
         0 &  $775.5 \pm 1$   & 791 &  $891.66 \pm 0.26$     &797&$1019.455 \pm 0.020 $   & 791& $782.65 \pm 0.12$   & 791    \\
\hline
         1 &  $1465 \pm 25$  &1570 &  $1414 \pm 15$  & 1573& $1680 \pm 20$    &1570 &$1400-1450$  &1570   \\
\hline
         2 & $1720 \pm 20$  & 1786  & $1717 \pm 27$    & 1788  & $2175 \pm 15$     &1786 &$1670 \pm 30$   &1786    \\
\hline
         3 & $1909 \pm 30$     & 1951  & ---                       &1955  & ---                  & 1951 & ---               &1951  \\
\hline
         4 & $2149 \pm 17$   & 2137  & ---                      & 2140 & ---                    &2137  & ---                &2137  \\
\hline
         5 & $2265 \pm 40$   &  2375   & ---                    &2377 & ---                    &2375    & ---                 & 2375  \\
\hline
\end{tabular}
\caption{The predicted mass spectra for vector mesons with including the instanton effects of determinant term.} \label{vectormassesk3}
\end{center}
\end{table}

 \begin{table}[ht!]
\begin{center}
\begin{tabular}{|c|c|c|c|c|c|c|c|c|}
\hline
        n &$a_1$ exp.(MeV)  &Theory & $K_1$ exp.(MeV)   &Theory & $f_1$ exp.(MeV)& Theory & $f_1$ exp.(MeV)& Theory   \\
\hline\hline
       0  &   $1230 \pm 40$    &1230 &  $1272 \pm 7$       &1385 &$1426.4 \pm 0.9$    &1447 &$1281.8 \pm 0.6$ &1347\\
\hline
         1 &  $1647 \pm 22$     &1683  &  $1403 \pm 7$    &1735 & ---                        &1758& $1518 \pm 5$   &1721 \\
\hline
         2 & $1930^{+30}_{-70}$  & 1879 & $1650 \pm 50$  &1923&---                         & 1942    &  ---           & 1911\\
\hline
         3 & $2096 \pm 122$      & 2043  &---                      &  2085&    ---                     &2104         & ---            & 2074\\
\hline
         4 & $2270^{+55}_{-40}$    &2222  & ---                   &2263 & ---                        &2281         &---             & 2252\\
\hline
         5&---&   2455                    &---& 2494            &---&2512               &---&   2484                      \\
\hline
\end{tabular}
\caption{The predicted mass spectra for axial-vector mesons with including the instanton effects of determinant term.} \label{axial-vectormassesk3}
\end{center}
\end{table}

It is seen that the instanton effects given by the determinant term improve the mass spectra for the groud state mesons, while the SU(3) flavor symmetry breaking effects remain small to explain all of the current experimental data.

\section{Meson Mixing Effects}\label{sec:mixing}

The non-diagonal elements of $M_A^2$, $M_S^2$, $\hat{M}_{A}^2$ and $\tilde{M}_S^2$ cause a mixing between SU(3) isosinglet and singlet states. All of the mixing effects arise from the difference between $v_s$ and $v_u$ due to the SU(3) flavor symmetry breaking. In this section, We shall discuss and evaluate the possible effects caused by the mixing terms which have been ignored in previous sections.

For a demonstration, let us begin with a simplified action:
\bea\label{mixaction}
 S&=&\int
d^5x\sqrt{g}e^{-\Phi(z)}\,{\rm{Tr}}\big{[}|\p \phi(z,x_i)|^2 + |\p \psi(z,x_i)|^2 + f_\phi(z)|\phi(z,x_i)|^2 +f_\phi(z)|\psi(z,x_i)|^2\nonumber\\
&& + 2f_{\phi\psi}(z)|\phi(z,x_i)\psi(z,x_i)|\big{]}
\eea
with $x_i$ denoting the 4-dimensional coordinators $x_0,x_1,x_2,x_3$.
After integrating over $z$, we then obtain an action at 4-dimensional spacetime:
\bea\label{mixaction4}
 S&=&\int
d^4x\sqrt{g}e^{-\Phi(z)}\,{\rm{Tr}}\big{[}|\p \phi_4(x_i)|^2 + |\p \psi_4(x_i)|^2 + m_\phi^2(x_i)|\phi_4(x_i)|^2 + m_\psi^2(x_i)|\psi_4(x_i)|^2\nonumber\\
&& + 2\Delta m_{\phi\psi}^2(x_i)|\phi_4(x_i)\psi_4(x_i)|\big{]}
\eea
with $\phi_4(x_i)=\sqrt{\int dz \sqrt{g}e^{-\Phi(z)} |\phi(z,x_i)|^2}$,
$m_\phi^2(x_i)=\frac{\int dz \sqrt{g}e^{-\Phi(z)} (f_\phi(z)|\phi(z,x_i)|^2+|g^{zz}\p_5\phi(z,x_i)\p_5\phi(z,x_i)|)}{\phi_4(x_i)^2}$ and other function
like $\psi_4(x_i)$ can be exported similarity. The terms $m_\phi^2$, $m_\psi^2$ and $\Delta m_{\phi\psi}^2$ are considered to form the mass square matrix for the fields $\phi$ and $\psi$:

\be
M^2=\left(
       \begin{array}{cc}
              m_\phi^2  &  \Delta m_{\phi\psi}^2\\
              \Delta m_{\phi\psi}^2  &  m_\psi^2\\
             \end{array}
           \right)
\ee
When it is a constant matrix, it can be diagonalized to obtain two independent mass eigenstates without mixing.

In the limit $f_{\phi\psi}(z)\to 0$, we can solve $\phi$ and $\psi$ independently with the solutions $\phi^0(z,x_i)=\phi^0_4(x_i)\phi^0_5(z)$ and $\psi^0(z,x_i)=\psi^0_4(x_i)\psi^0_5(z)$. Applying the similar operation for Eq.\,(\ref{mixaction}) and Eq.\,(\ref{mixaction4}), we arrive at:
\bea\label{mixaction40}
 S^0=\int
d^4x\sqrt{g}e^{-\Phi(z)}\,{\rm{Tr}}\big{[}|\p \phi^0_4(x_i)|^2 + |\p \psi^0_4(x_i)|^2
 + m_{0\phi}^2|\phi^0_4(x_i)|^2 + m_{0\psi}^2|\psi^0_4(x_i)|^2\big{]}
\eea

Thus in the case that $f_{\phi\psi}(z)$ is small enough, one can replace, as a good approximation, $\phi(z,x_i)$ and $\psi(z,x_i)$ in Eq.\,(\ref{mixaction}) by $\phi^0(z,x_i)$ and $\psi^0(z,x_i)$, and the mass matrix $M^2$ is given by $m_\phi^2=m_{0\phi}^2$, $m_\psi^2=m_{0\psi}^2$ and $\Delta m_{\phi\psi}^2= \int dz \sqrt{g}e^{-\Phi(z)} f_{\phi\psi}(z)|\phi_5^0(z)\psi_5^0(z)|$ (here $\phi$ and $\psi$ are normalized).

For the isosinglet and singlet scalar mesons, the effects of mixing part is given by
 \be\label{mixing}
\Delta m_{f_0}^2= \int _\epsilon^{z_m} dz M_{89}(z) S_8(z) S_9(z)
\ee
With $S_8(z)$ and $S_9(z)$ being the meson bulk wave functions for the ground states. Where $S_8(z)$ and $S_9(z)$ satisfy the normalization conditions:
    \be
   \int _\epsilon^{z_m}dz \,a(z)^3\,e^{-\Phi(z)}S_{n}(z)S_{m}(z)=\delta_{mn}
    \ee
and $M_{89}$ is given by
\be
M_{89}(z)=-a(z)^2\Big{\{}\frac{\lambda}{2\sqrt{2}}[v_s(z)^2-\vu(z)^2]+\frac{c_0}{6\sqrt{2}}[\vu(z)-v_s(z)]\Big{\}}.
\ee

For the isosinglet and singlet axial-vector mesons, the effects of mixing part have the same form as Eq.\,(\ref{mixing}), but with
a replacement
\be
M_{89}(z)=-(2\pi)^2a(z)^2\frac{\sqrt{2}}{3}((v_s(z)^2-\vu(z)^2)-c_1\frac{a(z)'}{a(z)^3}(v_s(z)v_s(z)'-\vu(z)\vu(z)'))
\ee
and the normalized axial-vector meson bulk wave functions $A_m(z)$ which satisfy the normalization conditions
\be
   \int _\epsilon^{z_m}dz \,a(z)\,e^{-\Phi(z)}A_{n}(z)A_{m}(z)=\delta_{mn}
    \ee
Here $z_m$ is chosen to be large enough to make the integration convergence. With above analysis, we can now make a calculation for the mixing effects. The numerical results are given in Table XIII without instanton effects of determinant term and Table XIV with instanton effects of determinant term. The mixing effects are found to be small in the present model.

\begin{table}[ht!]
\begin{center}
\begin{tabular}{|c|ccc|ccc|}
\hline
         $c_0=0$&$f_0(\text{GeV})$ & $f_0(\text{GeV})$ &$\Delta m_{f_0}^2 (\text{GeV}^2)$& $f_1(\text{GeV})$&$f_1(\text{GeV})$ & $\Delta m_{f_1}^2 (\text{GeV}^2)$\\
\hline\hline
experiment     &  0.980&0.600&-- &1.420 &1.285&-- \\
\hline\hline
mixed     &  0.433&0.374&-0.0335 &1.444 &1.338&-0.0742 \\
\hline
diagonalized   &  0.452& 0.350&-- &1.450 &1.331&-- \\
\hline
mixing angle(Degree) & &-3.10  & & & -0.261 &\\
\hline
\end{tabular}
\caption{Diagonalization of $f_0$ and $f_1$ with dilaton solution and input parameters given in section \ref{sec:includeS}.} \label{mixing part1}
\end{center}
\end{table}

 \begin{table}[ht!]
\begin{center}
\begin{tabular}{|c|ccc|ccc|}
\hline
        $c_0=2$&$f_0(\text{GeV})$ & $f_0(\text{GeV})$ &$\Delta m_{f_0}^2 (\text{GeV}^2)$& $f_1(\text{GeV})$&$f_1(\text{GeV})$ & $\Delta m_{f_1}^2 (\text{GeV}^2)$\\
\hline\hline
experiment     &  0.980&0.600&-- &1.420 &1.285&-- \\
\hline\hline
mixed     &  0.539&0.400&-0.0353 &1.447 &1.347&-0.0867\\
\hline
diagonalized        &  0.548& 0.389&-- &1.455&1.338&-- \\
\hline
mixing angle(Degree) & &-1.19  & & & -0.365&\\
\hline
\end{tabular}
\caption{Diagonalization of $f_0$ and $f_1$ with dilaton solution and input parameters given in section \ref{sec:determinant} with including the determinant term.} \label{mixing part2}
\end{center}
\end{table}

\section{Conclusions}\label{sec:conc}

We have investigated the infrared improved soft-wall AdS/QCD model with three flavors, and shown that the chiral $SU_L(3) \times SU_R(3)$ and $U_L(1) \times U_R(1)$ symmetry breaking and linear confinement can well be understood within such a simple model. The resulting resonance meson states agree well with the experimentally confirmed resonances. Except the quartic interaction term discussed in \cite{Sui:2009xe}, two additional quartic terms have been introduced to improve the mass spectra of SU(3) octet and singlet states, especially the ground state mesons. The special quartic term $(Tr\ln X-Tr\ln X^\dag)^2Tr(XX^\dag)$ which breaks chiral U(1) symmetry has been shown to interact with the singlet pseudoscalar when adding to the action, it then results in a better agreement for the prediction of $\eta'$ meson mass. We have also found that the quartic term $iA_{\mu\nu}(D^{\mu}X^{+}D^{\nu}X-D^{\nu}X^{+}D^{\mu}X)$ can improve the prediction for the ground state mass spectra of axial-vector mesons, and bring a better agreement with the experimental data. The instanton effects given by the determination term have also been discussed, though it can improve the prediction for the ground state mass spectra of scalar mesons, while its coupling coefficient cannot be too large, otherwise it may cause the instability of the dilaton solution and also the breaking of the GOMR relation for kaon meson. It is similar to the quartic term of meson field X, which may change the sign of the dilaton solution in the infrared region and destroy the special slope of dilaton needed to generate the linear confinement, so its coupling coefficient has to be set in an appropriate range.

We would like to point out that the simple predictive AdS/QCD model discussed in the present paper provides us an intuitive and also quantitative understanding on both the SU(3) chiral symmetry breaking and linear confinement, while it needs to be further improved and developed for a better understanding on the SU(3) flavor symmetry breaking in order to bring a more consistent prediction on the mass spectra of all SU(3) octet and singlet meson states. It would be interesting to further investigate possible contributions from other higher order terms. Again in the present considerations, the dilaton and gravity are treated as background fields, it would be important to study the dynamical features of dilaton field and consistently consider the 5D gravity effects from the
back-reacted geometry.

\section*{Acknowledgements}

This work was supported in part by the National Science Foundation of China (NSFC) under the grant \# 10821504, 10975170 and the key Project of Knowledge Innovation Program (PKIP) of Chinese Academy of Science.



\end{document}